\documentclass[12pt,a4paper]{article}
\usepackage[usenames]{color}
\usepackage{amsfonts, amssymb, amsmath, amsthm, latexsym,amscd}
\usepackage[dvips,matrix,ps,color,line,graph]{xy}

\begin{document}

\title{ Structure, Individuality and Quantum Gravity}
\author{John Stachel\thanks{Department of Physics and Center for Einstein Studies, 
Boston University; {\tt john.stachel@gmail.com} }}
\date{}
\maketitle

\begin{abstract}

After reviewing various interpretations of {\it structural realism}, I adopt here a definition that allows both relations between things that are already individuated (which I call ``relations between things'') and relations that individuate previously un-individuated entities ("things between relations").  Since both space-time points in general relativity and elementary particles in quantum theory fall into the latter category, I propose a principle of maximal permutability  as a criterion for the fundamental entities of any future theory of ``quantum gravity''; i.e., a theory yielding both general relativity and quantum field theory in appropriate limits. Then I review of a number of current candidates for such a theory. First I look at the effective field theory and asymptotic quantization approaches to general relativity, and then at string theory. Then a discussion of some issues common to all approaches to quantum gravity based on the full general theory of relativity argues that  processes, rather than states should be taken as fundamental in any such theory. A brief discussion of the canonical approach is followed by a survey of causal set theory, and a new approach to the question of which space-time structures should be quantized ends the paper.

\end{abstract}
\pagebreak

\tableofcontents
\pagebreak
\section{What is Structural Realism?}

The term``structural realism'' can be (and has been) interpreted in a number of different ways.\footnote{For a recent survey, with references to earlier literature, see the symposium "Structural Realism and Quantum Field Theory," \cite{Symons2003}, which includes papers by Tian Yu Cao, Steven French and James Ladyman, and Simon Saunders.}  I assume that, in discussions of structuralism, the concept of  "structure" refers to some set of relations between the things or entities that they relate, called \textit{the relata}. Here I interpret "things" in the broadest possible sense: they may be material objects, physical fields,  mathematical concepts, social relations, processes, etc. \footnote{In this section, "thing" is used in a sense that includes processes. Note that from {\it Section 5a} on, it is used in a more restricted sense, in which ``thing'' is contrasted with "process."} People have used the term ``structural realism'' to discribe different approches to the nature of the relation between things and relations. These differences all seem to be variants of three basic possibilities:\\

{\bf I. There are only  relations without relata.}\footnote{See \cite{Krause2004}, pp. 1, for the phrase ``relations without the relata''}  \\

As applied to a particular relation, this assertion seems incoherent. It only makes sense if it is interpreted as the metaphysical claim that ultimately there are only relations; that is, in any given relation, all of its {\it relata} can in turn be interpreted as {\it relations}. Thus, the totality of structural relations reduces to relations between relations between relations . As Simon Saunders might put it, it's relations all the way down. \footnote{``I believe that objects are structures; I see no reason to suppose that there are ultimate constituents of the world, which are not themselves to be understood in structural terms. So far as I am concerned, it is turtles all the way down" (\cite{Saunders2003}, pp. 329).}
It is certainly true that, in certain cases, the relata can themselves be interpreted as relations; but I would not want to be bound by the claim that this is always the case. 
I find rather more attractive the following two possibilities: \\

{\bf II. There are relations, in which the things are primary and their relation is secondary.}\\

{\bf III. There are  relations, in which the relation is primary while the things are secondary.}\\

In order to make sense of either of these possibilities, and hence of the distinction between them, one must assume that there is always a distinction between the {\it essential} and {\it non-essential} properties of any thing,\footnote{For example, in quantum mechanics, electrons are characterized by their essential properties of mass, spin and charge. All other properties that they may exhibit in various  processes - such as positions, momenta, or energies -  are non-essential (see note {\tt 20}). For a discussion of their haecceity, see the next section). As this example suggests, the distinction between essential and non-essential properties - and indeed the distinction between elementary and composite entities - may be theory-dependent (see \cite{Dosch2004}).  (Note that having the same essence is what characterizes a natural kind.)}  For {\bf II} to hold (i.e. things are primary and their relation is secondary),  no essential property of the relata can depend on the particular relation under consideration; while for {\bf III} to hold( i.e. the relation is primary and the relata are secondary ), at least one essential property of each of the relata must depend on the relation. Terminology differs, but one widespread usage denotes relations of type {\bf II} as external, those of type {\bf III} as internal. 
One could convert either possibility into a metaphysical doctrine: ``All relations are external'' or ``All relations are internal''; and some philosophers have done so. But, in contradistinction to {\bf I}, there is no need to do so to make sense of {\bf II} and {\bf III}. If one does not, then the two are perfectly compatible.

Logically, there is a fourth possible case:\\

{\bf IV. There are things, such that any relation between them is only apparent. }\\
 
 This is certainly possible in particular situations. One could, for example, pre-program two mechanical dolls (the things) so that each would move independently of the other, but in such a way that they seemed to be dancing with each other (the apparent relation - I assume that ``dancing together'' is a real relation between two people). 
 
Again, one might convert this possibility into a universal claim: ``All relations are only apparent.''. Leibniz' monadology, for example, might be interpreted as asserting that all relations between monads are only apparent. Since God set up a pre-established harmony among them, they are pre-programmed to behave as if they were related to each other. As a metaphysical doctrine, I find {\bf IV} even less attractive than {\bf I}. And if adopted,  it could hardly qualify as a variant of structural realism, so I shall not mention {\bf IV} any further.

While several eminent philosophers of science (e.g. French and Ladyman) have opted for version {\bf I} of structural realism, to me versions {\bf II} and {\bf III} (interpreted non-metaphysically) are the most attractive. They do not require commitment to any metaphysical doctrine, but allow for a decision on the character of the relations constituting a particular structure on a case-by-case basis. \footnote{For further discussion of cases  {\bf II} and  {\bf III}, see \cite{Stachel2002}, which refers to case  {\bf II} as ``relations between things,'' and to case  {\bf III} as ``things between relations.''}
My approach leads to a picture of the world, in which there are entities of many different natural kinds, and it is inherent in the nature of each kind to be structured in various ways. These structures themselves are organized into various structural hierarchies, which do not all form a linear sequence (chain); rather, the result is something like a totally partially-ordered set of structures. This picture is dynamic in two senses: there are changes in the world, and there are changes in our knowledge of the world. 
 
As well as a {\it synchronic} aspect, the entities and structures making up our current picture of the world have a {\it diachronic} aspect: they arise, evolve, and ultimately disappear - in short, they constitute {\it processes}. And our current picture is itself subject to change. What particular entities and structures are posited, and whether a given entity is to be regarded as a thing or a relation, are not decisions that are forever fixed and unalterable; they may change with changes in our empirical knowledge and/or our theoretical understanding of the world. So I might best describe this viewpoint as a dynamic structural realism. \footnote{For further discussion of the structural hierarchy, see \cite{Stachel2004a}. For many examples of such hierarchies in physics, biology and cosmology, see \cite{Ellis2002}. Although my concepts of entity and structure are meant to be ontological, the term  ``ontic structural realism'' has been preempted and given a different significance (see \cite{Ladyman1998}).} 

\section{Structure and Individuality} 

A more detailed discussion of many points in this section is presented in \cite{Stachel2004a}, \cite{Stachel2005}.

It seems that, as deeper and deeper levels of  these structural hierarchies are probed, the property of inherent individuality that characterizes more complex, higher-level entities
- such as a particular crystal in physics, or a particular cell in biology– is lost. Using some old philosophical terminology, I say that a level at has been reached, which the entities characterizing this level possess quiddity  but not haecceity. ``Quiddity'' refers to the essential nature of an entity, its natural kind; and-- at least at the deepest level which we have reached so far - entities of different natural kinds exist, e.g., electrons, quarks, gluons, photons, etc. \footnote{Believers in a unified ``Theory of Everything'' will hope that ultimately only entities of one natural kind will be needed, and that all apparently different kinds will emerge from the relational properties of the one fundamental quiddity. String theory might be regarded as an example of such a theory; but, aside from other problems, its current framework is based on a fixed a background space-time, as will be discussed in {\it section 3}.} What distinguishes entities of the same natural kind (quiddity) from each other, their unique individuality or ``primitive thisness,'' is called their ``haecceity.''\footnote{Runes 1962: ``Haecceity..[is].. A term employed by Duns Scotus to express that by which a quiddity, or general essence, becomes an individual, particular nature, or being.'' \cite{Teller1995} (see pp. 17, note 2), following \cite{Adams1979}, noted the utility of the term "haecceity,"and his suggestion has been followed by many philosophers of physics.}  Traditionally, it was always assumed that every entity has such a unique individuality: a haecceity as well as a quiddity.  However, modern physics has reached a point, at which we are led to postulate entities that have quiddity but no haecceity that is inherent, i.e., independent of the relational structures in which they may occur. In so far as they have any haecceity (and it appears that degrees of haecceity must be distinguished\footnote{ For example, the electrons confined to a particular ``box'' (i.e., infinite potential well) may be distinguished from all other electrons, if not from each other.}), such entities inherit it from the structure of relations in which they are enmeshed. In this sense, they are indeed examples of the case {\bf III}: ``things between relations.''\cite{Stachel2002}

Since Kant, philosophers have often used position in space as a principle of individuation for otherwise indistinguishable entities;  more recently, similar attempts have been made to individuate physical events or processes \footnote{See, e.g., \cite{Auyang1995}, Chapter 6}.

A physical process occupies a (generally finite) region of space-time; a physical event is supposed to occupy a point of space-time. In theories, in which space-time is represented by a continuum, an event can be thought of as the limit of a portion of some physical process as all the dimensions of the region of space-time occupied by this portion are shrunk to zero. Classically, such a limit may be regarded as physically possible, or just as an ideal limit. ``An event may be thought of as the smallest part of a process .... But do not think of an event as a change happening to an otherwise static object. It is just a change, no more than that'' (\cite{Smolin2002a}, pp. 53). See section 5.1 for further discussion of processes. It is probably better to avoid attributing physical significance to point events, and accordingly to mathematically reformulate general relativity in terms of sheaves\footnote{See  
\cite{Stachel2005} for further discussion of this point. For one such reformulation of differential geometry, see \cite{Mallios1998}, and for applications to general relativity, see \cite{Mallios2004} and \cite{Mallios2003}.}

Individuation by means of position in space-time works at the level of theories with a fixed space-time structure, notably special-relativistic theories of matter and/or fields\footnote{Actually, the story is more complicated than this. The points of Minkowski space-time, for example, are themselves homogeneous, and some physical framework must be introduced in order to physically individuate them. Only after this has been done, can these points be used to individuate other events or processes. The physical framework may be fixed non-dynamically (e.g., by using rods and clocks introduced a priori); or if fixed by dynamical process (e.g., light rays and massive particles obeying dynamical equations), the resulting individuation must be the same for all possible dynamical processes. This is better said in the language of fiber bundles, in which particular dynamical physical fields are represented by cross-sections of the appropriate bundle: If the metric of the base space is given a priori, the individuation of the points of the base space is either also so given, or is the same for all cross-sections of the bundle (see \cite{Stachel2005}).} but, according to general relativity, because of the dynamical nature of all space-time structures, \footnote{Except for continuity, differentiability and local topology of the differentiable manifold.} the points of space-time lack inherent haecceity; thus they cannot be used for individuation of other physical events in a general-relativistic theory of matter and/or non-gravitational fields. This is the purport of the ``hole argument'' (see \cite{Stachel1993} and earlier references therein). The points of space-time have quiddity as such, but only gain haecceity (to the extent that they do) from the properties they inherit from the metrical or other physical relations imposed on them.
\footnote{This is better said in the language of fiber bundles, in which particular dynamical physical fields are represented by cross-sections of the appropriate bundle: If the metric of the base space is given a {\it priori}, the individuation of the points of the base space must also be given a priori, or in such a way as to be the same for all cross-sections of the bundle (see \cite{Stachel2005}).} 
In particular, the points can obtain haecceity from the inertio-gravitational field associated with the metric tensor: For example, the four non-vanishing invariants of the Riemann tensor in an empty space-time can be used to individuate these points in the generic case (see ibid., pp. 142-143)\footnote{Again this is better said in the language of fibered manifolds, in which particular dynamical physical fields are represented by cross-sections of the manifold: One can now define the base space as the quotient of the total space by the fibration. Thus, even the points of the base space (let alone its metric) are not defined {\it a priori}, and their individuation depends on the choice of a cross-section of the fibered manifold, which will include specification of a particular inertio-gravitational field. For a detailed discussion, see \cite{Stachel2005}.}

Indeed, as a consequence of this circumstance, in general relativity the converse attempt has been made: to individuate the points of space-time by means of the individuation of the physical (matter or field) events or processes occurring at them; i.e., by the relation between these points and some individuating properties of matter and/or non-gravitational fields. Such attempts can succeed at the macroscopic, classical level; but, if the analysis of matter and fields is carried down far enough - say to the level of the sub-nuclear particles and field quanta\footnote{I reserve the term ``elementary particles'' for fermions and ``field quanta'' for bosons, although both are treated as field quanta in quantum field theory. I aim thereby to recall the important difference between the two in the classical limit: classical particles for fermions and classical fields for bosons. In this paper, I am sidestepping the question of whether and when the field concept is more fundamental than the particle concept in quantum field theory, especially in non-flat background space-times (I will discuss it in detail in 
\cite{Stachel2004b}, but see the next note and note 40); but in the special-relativistic theories, a preparation or registration may involve either gauge-invariant field quantities or particle numbers.} -  then the particles and field quanta of differing quiddity all lack inherent haecceity.\footnote{At level of non-relativistic quantum mechanics for a system consisting of a fixed number of particles of the same type, this is seen in the need to take into account the bosonic or fermionic nature of the particle in question by the appropriate symmetrization or antisymmetrization procedure on the product of the one-particle Hilbert space (see, e.g., 
\cite{Haag1996}, pp. 35-36, for more details). At the level of special-relativistic quantum field theory, in which interactions may change particle numbers, it is seen in the notion of field quanta, represented by occupation numbers (arbitrary for bosons, either zero or one for fermions) in the appropriately constructed Fock space; these quanta clearly lack individuality. (See, e.g.,  \cite{Teller1995}, \cite{Haag1996}, pp. 36-38.). At the level of quantum field theory in background curved space-times, "a useful particle interpretation of states does not, in general, exist" (\cite{Wald1994}, p. 47).}  Like the points of space-time, insofar as they have any individuality, it is inherited from the structure of relations in which these quanta are embedded. For example, in a process involving a beam of electrons, a particular electron may be individuated by the click of a particle counter.\footnote{The macroscopic counter is assumed to be inherently individuated. It seems that, for such individuation of an object, a level of structural complexity must be reached, at which it can be uniquely and irreversibly ``marked'' in a way that distinguishes it from other objects of the same nature (quiddity). My argument is based on an approach, according to which quantum mechanics does not deal with quantum systems in isolation, but only with processes that such a system can undergo. (For further discussion of this approach, see \cite{Stachel1986}, \cite{Stachel1997}). A process  (Feynman uses "process," but Bohr uses ``phenomenon'' to describe the same thing) starts withs the preparation of the system, which then undergoes some interaction(s), and ends with the registration of some result (a ``measurement''). In this approach, a quantum system is defined by certain essential properties (its quiddity); but manifests other, non-essential properties (its haecceity) only at the beginning (preparation) and end (registration) of some process. (Note that the initially-prepared properties need not be the same as the finally-registered ones.) The basic task of quantum mechanics is to calculate a probability amplitude for the process leading from the initially prepared-values to the finally-registered ones. (I assume a maximal preparation and registration – the complications of the non-maximal cases are easily handled). (See Section 5a  for a discussion of whether this interpretation of quantum mechanics is viable in the context of quantum gravity.\cite{Wald1994}, \cite{Wald})}

 In all three of these cases - space-time points or regions in general relativity,  elementary particles in quantum mechanics, and field quanta in quantum field theory -  insofar as the fundamental entities have  haecceity, they inherit it from the structure of relations in which they are enmeshed. But there is an important distinction here between general relativity one the one hand and quantum mechanics and quantum field theory on the other: the former is background-independent while the latter are not; but I postpone further discussion of this difference until Section 5b.
 
What has all this to do with the search for a theory of quantum gravity? The theory that we are looking for must underlie both classical general relativity and quantum theory, in the sense that each of these two theory should emerge from ``quantum gravity'' by some appropriate limiting process. Whatever the ultimate nature(s) (quiddity) of the fundamental entities of a quantum gravity theory turn out to be, it is hard to believe that they will possess an inherent individuality (haecceity) already absent at the levels of both general relativity and quantum theory (see \cite{Stachel2004a}). So I am led to assume that, whatever the nature(s) of the fundamental entities of quantum gravity, they will lack inherent haecceity, and that such individuality as they manifest will be the result of the structure of dynamical relations in which they are enmeshed.
Given some physical theory, how can one implement this requirement of no inherent haecceity? Generalizing from the previous examples, I maintain that the way to assure the inherent indistinguishability in of the fundamental entities of the theory is to require the theory to be formulated in such a way that physical results are invariant under all possible permutations of the  basic entities of the same kind (same quiddity).\footnote{In this principle, the word "possible" is to be understood in the following sense: If the theory is formulated in such a way that some dynamically-independent permutations of the fundamental entities are possible, then the theory must be invariant under all permutations. Or the theory must be formulated in such a way that no such permutations are possible (see note 16)}  I have named this requirement the {\it principle of maximal permutability}. (See \cite{Stachel2005} for a more mathematically detailed discussion.)  
 
The exact content of the principle depends on the nature of the fundamental entities. For theories , such as non-relativistic quantum mechanics, that are based on a finite number of discrete fundamental entities, the permutations will also be finite in number, and maximal permutability becomes invariance under the full symmetric group. For theories, such as general relativity, that are based on fundamental entities that are continuously, and even differentiably related to each other, so that they form a differentiable manifold, permutations become diffeomorphisms. For a diffeomorphism of a manifold is nothing but a continuous and differentiable permutation of the points of that manifold.\footnote{Here, diffeomorphisms are to be understood in the active sense, as point transformations acting on the points of the manifold, as opposed to the passive sense, in which they act upon the coordinates of the points, leading to coordinate re-descriptions of the same point. See \cite{Stachel2005} for a more detailed discussion, based on the use of fibered manifolds and local diffeomorphisms.}  So, maximal permutability becomes invariance under the full diffeomorphism group. Further extensions to an infinite number of discrete entities or mixed cases of discrete-continuous entities, if needed, are obviously possible.

In both the case of non-relativistic quantum mechanics and of general relativity, it is only through dynamical considerations that individuation is effected. In the first case, it is through specification of a possible quantum-mechanical process that the otherwise indistinguishable particles are individuated (``The electron that was emitted by this source at 11:00 a.m. and produced a click of that Geiger counter at 11:01 a.m.''). In the second case, it is through specification of a particular solution to the gravitational field equations that the points of the space-time manifold are individuated (``The point at which the four non-vanishing invariants of the Riemann tensor had the following values: ...''). So one would expect the principle of maximal permutability of the fundamental entities of any theory of quantum gravity to be part of a theory in which these entities are only individuated dynamically. 

Thomas Thiemann has pointed out that, in the passage from classical to quantum gravity, there is good reason to expect diffeomorphism invariance to be replaced by some discrete combinatorial principle:
 \begin{quote}
The concept of a smooth space-time should not have any meaning in a quantum theory of the gravitational field where probing distances beyond the Planck length must result in black hole creation which then evaporate in Planck time, that is, spacetime should be fundamentally discrete. But clearly smooth diffeomorphisms have no room in such a discrete spacetime. The fundamental symmetry is probably something else, maybe a combinatorial one, that looks like a diffeomorphism group at large scales. (\cite{Thiemann2001}, pp. 117)
\end{quote}

In the next section, I shall look at the effective field theory approach to general relativity and asymptotic quantization, and then, in the following section, at string theory, both in the light of the principle of maximal permutability. Section 5 discusses some issues common to all general-relativity-based approaches to quantum gravity. I had hoped to treat loop quantum gravity in detail in this paper, but the discussion outgrew my allotted spatial bounds; so just a few points about the canonical approach are discussed in Section 6, and the fuller discussion relegated to a separate paper, \cite{Stachel2004b}. Section 7 is devoted to causal set theory, and Section 8 sketches a possible new approach, suggested by causal set theory, to the question of what space-time structures to quantize.

\section{Effective field theory approach and asymptotic quantization}

The earliest attempts to quantize the field equations of general relativity were based on treating it using the methods of special-relativistic quantum field theory, perturbatively expanding the gravitational field around the fixed background Minkowski space metric and quantizing only the perturbations. By the 1970s, the first wave of such attempts petered out with the realization that the resulting quantum theory is perturbatively non-renormalizable. With the advent of the effective field theory approach to non-renormalizable quantum field theories, a second, smaller wave arose\footnote{Presumably (to continue the metaphor), the fashion for string theory pre-empted the waters, in which the second wave would otherwise have flourished. Of course, the effective field theory approach has been applied effectively to perturbatively renormalizable theories (see \cite{Dosch2004})} with the more modest aim of developing an effective field theory of quantum gravity valid for sufficiently low energies (for reviews, see \cite{Donoghue1995}, \cite{Burgess2004}). As is the case for all effective field theories, this approach is not meant to prejudge the nature of the ultimate resolution of ``the more fundamental issues of quantum gravity'' (
\cite{Burgess2004}, pp. 6), but to establish low-energy results that will be reliable whatever the nature of the ultimate theory.\footnote{``Effective field theory is to gravitation as chiral perturbation theory is to quantum chromodynamics – appropriate at large distances, and impotent at short,'' James Bjorken, ``Preface'' to \cite{Rovelli2004}, pp. xiii.}

The standard accounts of the effective field approach to general relativity take the metric tensor as the basic field, which somewhat obscures the analogy with Yang-Mills fields:

\begin{quote}
Despite the similarity to the construction of the field strength tensor of Yang Mills field theory, there is the important difference that the [Riemannian] curvatures involve two derivatives of the basic field, $R\sim \partial \partial g$. (\cite{Donoghue1995}, pp. 4)
\end{quote}

But much of the recent progress in bringing general relativity closer to other gauge field theories, and in developing background-independent quantization techniques, has come from giving equal importance (or even primacy) to the affine connection as compared to the metric (see Sections 6, 8 and \cite{Stachel2004b}). Since the curvature tensor involves only one derivative of the connection,  $R\sim \partial \Gamma$,  this approach brings the formalism of general relativity much closer to the gauge approach used treat all other interactions. From this point of view, one role of the metric tensor is to act as potentials for the connection     $\Gamma \sim \partial g$. From this viewpoint, one can reformulate the starting point of general relativity as follows. 
 
 The equivalence principle demands that inertia and gravitation be treated as intrinsically united, the resulting inertio-gravitational field being represented mathematically by a non-flat affine connection $\Gamma$ \footnote{See \cite{Stachel2006a} for a detailed discussion of this approach. Note that in this brief account I represent each geometric object by a single symbol, omitting all indices.}.  

If one assumes that this connection is metric, i.e., that the connection can be derived from a second-rank covariant metric field $g$ then according to general relativity such a non-flat metric field represents the chrono-geometry of space-time.
 
But the effective field approach assumes that the true chrono-geometry of space-time remains the Minkowski space-time of special relativity, represented by the fixed background metric $\eta$ \footnote{Of course, one could start with any fixed background metric—flat or non-flat-- and perform such a split between the inertial connection associated with this metric and the gravitational field tensor, the latter again being the difference between the inertio-gravitational connection associated with the g-field and  the inertial connection associated with the background metric. But regardless of space-time metric chosen, the kinematics of the theory is based on this fixed background metric; and this limits the space-time diffeomorphisms, under which the theory is presumed invariant, to the symmetries (if any) of this metric.}.  There is an unique, flat 
affine connection $\{ \}$ compatible with the Minkowski metric $\eta$ \footnote{I use this symbol as a reminder that the Christoffel symbols define this connection.},  and since  the difference between any two connections is a tensor, $\Gamma -\{ \}$ -  the difference between the non-flat and flat connections - is a tensor that serves to represents a purely gravitational field. 

Thus, the upshot of this approach is to violate the purport of the equivalence principle, according to which inertia and gravitation are essentially the same and should remain inseparable. With the help of the flat background metric and connection, they have been separated; and a kinematics introduced based on the purely inertial connection, a kinematics that is independent of the dynamics embodied in the purely gravitational tensor. 
The background metric is assumed to be unobservable, because the effect of the gravitational field on all (ideal) rods and clocks is to distort their measurements in such a way that they always map out the non-flat chrono-geometry that can be associated with the metric of the $g$-field. If effective field theory did not tell us better, we might be tempted to think of  this metric as the true chrono-geometry — but then we would be doing general relativity

Contrary to the purport of the equivalence principle, inertia and gravitation have been separated with the help of the background metric and a kinematics based on the background fields has been introduced that is independent of the dynamics of this gravitational tensor. However, the background metric is unobservable: The effect of this gravitational field on all (ideal) rods and clocks is to distort their measurements in such a way that they map out the non-flat chrono-geometry associated with the $g$-field, which if we did not know better, we would be tempted to think of as the true chrono-geometry \footnote{The reader may be reminded of the Lorentzian approach to special relativity: There really is a privileged inertial frame ("the ether frame"). But motion through the ether contracts all (ideal) measuring rods, slows down all (ideal) clocks, and increases all masses in such a way that this motion is undetectable. Clocks in a moving inertial frame synchronized by light signals, but  forgetting about the effects of this motion on the propagation of light, will read the ``local time'' and light will appear to propagate with the same speed c in all inertial frames I often wonder why those who adopt the special-relativistic approach to general relativity don't abandon special relativity too in favor of the ether frame? } The points of the background metric (flat or non-flat 
- see note 26) are then assumed to be individuated up to the symmetry group of this metric, which at most can be a finite-parameter Lie group (e.g., the ten parameter Poincaré group for the Minkowski background metric) acting on the points of space-time.\footnote{We merely mention the additional global complications: The topology chosen for the background space-time (e.g., $\mathrm{R}^4$ for Minkowski space-time) may not be the same as the topology associated with particular g-fields that solve the field equations (e.g., the Kruskal manifold for the Schwarzschild metric). As Trautman showed long ago, if one solves for the g-field of a point mass at the origin by successive approximations, at each order the solution field is regular except at the origin. But when the infinite series is summed, the Schwarschild solution is obtained with its quite different topology.}  

Since the full diffeomorphism group acting on the base manifold is not a symmetry group of the background metric, \footnote{This symmetry group includes only those diffeomorphisms generated by the Killing vectors of the background metric.} this version of quantum gravity does not meet our criterion of maximal permutability. If we choose a background space-time with no symmetry group, each and every point of the background space-time manifold will be individuated by the non-vanishing invariants of the Riemann tensor. But if there is a symmetry group generated by one or more Killing vectors, then points on the orbits of the symmetry group will not be  so individuated, but must be individuated by some additional non-dynamical method.\footnote{In the extreme case of Minkowski space time, for which all the invariants of the Riemann tensor vanish, the individuation of all the points must be non-dynamical. However, the situation is not as bad as this comment might suggest. Singlng out of one point as origin, of three orthogonal directions, of a unit of spatial distance and a unit of time serve to complete the individuation of all remaining points.}

Other diffeomorphisms can only be interpreted {\it passively}, as coordinate redescriptions of the background space-time and inertial fields. They can be given an {\it active} interpretation only as {\it gauge transformations} on the gravitation potentials $h=g-\eta$
\footnote{That is, such a transformation can either be looked upon either passively, as a transformation to a non-inertial frame of reference, in which inertial forces appear; or actively as a transformation of the gravitational field, producing additional gravitational forces. Of course, it can be divided into two parts, each of which is given a different one of the two interpretations. When the gravitational field is quantized and the background field is not, the need for a choice makes it evident that this interpretation of general relativity differs from the standard, diffeomorphism-invariant interpretation in more than matters of taste.}
 
Since the effective field approach does not claim to be any more than a low-energy approximation to any ultimate theory of quantum gravity, rather than an obstacle to any theory making such a claim, this approach presents a challenge. Can such a theory demonstrate that, in an appropriate low-energy limit, its predictions match the predictions of the effective field theory for experimental results? \footnote{Since the effective field theory approach is based on the assumption of a space-time continuum, the challenge will take a radically different form depending on whether the more fundamental theory is itself based on the continuum concept, as is string theory; or whether it denies basic significance to the space-time continuum, as do loop quantum gravity and causal set theory.} Since these experimental predictions will essentially concern low energy scattering experiments involving gravitons, it will be a long time indeed before any of these predictions can be compared with actual experimental result; and the effective field theory approach has little to offer in the way of predictions for the kind of experimental results that work on phenomenological quantum gravity 
is actually likely to give us in the near future.
 
In a sense, one quantum gravity program has already met this challenge: Ashtekar's (1987) asymptotic quantization, in which only the gravitational in - and out- fields at null infinity - i.e., at $\Im +$ (scri-plus) and $\Im -$ (scri-minus) - are quantized. Without the introduction of any background metric field, it is shown how non-linear gravitons may be rigorously defined in terms of these fields as irreducible representations of the symmetry group at null infinity. This group, however, is not the Poincar\'{e} group at null infinity, but the much larger Bondi-Metzner-Sachs group, which includes the super-translations depending on functions of two variables rather than the four paremeters of the translation group. This group defines a unique kinematics at null infinity that is independent of the dynamical degrees of freedom, and it is this decoupling of kinematics and dynamics that enables the application of more-or-less standard quantization techniques. Just as the quotient of the Poincaré group by its translation subgroup defines the Lorentz group, so does the quotient of the B-M-S group by its super-translation subgroup. Since, in both the effective field   and asymptotic quantization techniques, experiments in which the graviton concept could be usefully invoked involve the preparation of in-states and the registration of out states, there must be a close relation between the two approaches; although, as far as I know, this relation has not yet been elucidated in detail.

In summary, both the effective field theory and asymptotic quantization approaches avoid the difficulties outlined in the previous section by separating out a kinematics that is independent of dynamics. In the former case, this separation is {\it imposed by flat} everywhere on the space-time manifold by singling out a background space-time metric and corresponding inertial field, with the expectation that the results achieved will always be valid to good approximation in the low-energy limit of general relativity.
In the latter case, the separation is achieved only for the class of solutions that are asymptotically flat at null infinity (or more explicitly, the Riemann tensors of which vanish sufficiently rapidly in all null directions to allow the definition of null infinity). It is then proved that at at null infinity a kinematics can be decoupled from the dynamics at null infinity due to the symmetries of any gravitational field there, and that this can be done  without violating diffeomorphism invariance in the interior region of space-time. Again, this approach presents a challenge to any background-independent quantization program: derive the results of the asymptotic quantization program from the full quantum gravity theory in the appropriate limit.
 
\section{String Theory}

String (or superstring) theory applies the methods of special-relativistic quantum theory to two-dimensional time-like world sheets, called "strings." \footnote{"String theory is an ordinary QFT but not in the usual sense. It is an ordinary scalar QFT on a 2d Minkowski space, however, the scalar fields themselves are coordinates of the ambient target Minkowski space which in this case is 10 dimensional. Thus, it is similar to a first quantized theory of point particles" (\cite{Thiemann2002}, p. 12).} All known (and some unknown) particles and their interactions, including the graviton and the gravitational interaction,  are supposed to emerge as certain modes of excitation of and interactions between quantized strings. The fundamental entities of the original (perturbative) string theory are the strings– two-dimensional time-like world sheets - embedded in a given background space-time, the metric of which is needed to formulate the action principle for the strings. For that reason, the theory is said to be ``background-dependent.'' Quantization of the theory requires the background space-time to be of ten or more dimensions. \footnote{However, the so-called Pohlmayer string can be quantized in any number of dimensions, including four dimensionsal Minkowski space (see \cite{Thiemann2004}).}
 
The theory is seen immediately to  fail the test of maximal permutability since the strings are assumed to move around and vibrate in this background, non-dynamical space-time. So the background space-time, one of the fundamental constituents of the theory; is invariant only under a finite-parameter Lie subgroup (the symmetry group of this space-time, usually assumed to have a flat metric with Lorentzian signature) of the group of all possible diffeomorphisms of its elements. Many string theorists, with a background predominantly in special-relativistic quantum field theory (attitudes are also seen to be background-dependent), initially found it difficult to accept such criticisms; so it is encouraging that this point now seems to be widely acknowledged in the string community. \footnote{See \cite{Stachel2003}, pp. 31-32 for quotations to this effect from review articles by Michael Green and Thomas Banks.}
String theorist Brian Greene, recently presented an appealing vision of what a string theory without a background space-time might look like, but emphasized how far string theorists still are from realizing this vision:
Since we speak of the ``fabric'' of spacetime, maybe spacetime is stitched out of strings much as a shirt is stitched out of thread. That is, much as joining numerous threads together in an appropriate pattern produces a shirt's fabric, maybe joining numerous strings together in an appropriate pattern produces what we commonly call spacetime's fabric. Matter, like you and me, would then amount to additional agglomerations of vibrating strings– like sonorous music played over a muted din, or an elaborate pattern embroidered on a plain piece of material– moving within the context stitched together by the strings of spacetime. ....[A]s yet no one has turned these words into a precise mathematical statement. As far as I can tell, the obstacles to doing so are far from trifling. .... We [currently] picture strings as vibrating in space and through time, but without the spacetime fabric that the strings are themselves imagined to yield through their orderly union, there is no space or time. In this proposal, the concepts of space ad time fail to have meaning until innumerable strings weave together to produce them.
 
Thus, to make sense of this proposal, we would need a framework for describing strings that does not assume from the get-go that they are vibrating in a preexisting spacetime. We would need a fully spaceless and timeless formulation of string theory, in which spacetime emerges from the collective behavior of strings.
Although there has been progress toward this goal, no one has yet come up with such a spaceless and timeless formulation of string theory– something that physicists call a background-independent formulation [of the theory] ... Instead, all current approaches envision strings as moving and vibrating through a space-time that is inserted into the theory "by hand"... Many researchers consider the development of a background-independent formulation to be the single greatest unsolved problem facing string theory (\cite{Greene2004}, pp. 487-488).
One of the main goals of the currently sought-for M-theory (see \cite{Greene2004}, Chap. 13, pp. 376-412)  is to overcome this defect, but so far this goal has not been reached.

\section{Quantum general relativity - some preliminary problems}

String theory attempts to produce a theory of everything, including a quantum theory of gravity that will have general relativity (or a reasonable facsimile thereof) as part of its classical limit. Most other approaches to quantum gravity start from classical general relativity. In this section, I shall discuss two related issues that arise in the course of any such attempt.

\subsection{States or Processes: Which is primary ?}

There has been a long-standing debate between adherents of covariant and canonical approaches to quantum gravity. The former attempt to develop a four-dimensionally-invariant theory of quantum gravity from the outset; the latter start from a (3+1)-breakup of space-time, emphasizing three-dimensional spatial invariance, developing quantum kinematics before quantum dynamics. Christian W\"{u}thrich has related this debate to the philosophical debate between proponents of the endurance view of time and those of the perdurance view [which] reflects a disagreement concerning whether, and to what degree, time is on a par with spatial dimensions.  
According to the former view, ''an object is said to endure just in case it exists at more than one time.'' According to the latter view, 
objects perdure by having different temporal parts at different times with no part being present at more than one time. Perdurance implies that two [space-like] hypersurfaces ... do not share enduring objects but rather harbour different parts of the same four-dimensional object.\cite{Wuthrich2003}

I shall use slightly different terminology to make this important distinction. One approach to the quantum gravity problems places primary emphasis on the three-dimensional state of some thing; from this point of view, a process is just a succession of different states of this  thing. (The relation of this succession of states to some concept of  ``time'' is a contentious issue). The other approach places primary emphasis on four-dimensional processes; from this point of view, a "state" is just a particular spatial cross-section of a process and of secondary importance: all such cross-sections are equal,  and each sequence of states represents a different ``perspective'' on the same  process.
 
In pre-relativistic physics, the absolute time provided a natural foliation of space-time into spatial cross-sections. So, even if one favored the ``process'' viewpoint for philosophical reasons, there was little harm to physics - if not to philosophy - in using the alternate "state" viewpoint. While the split into spaces was not unique (one inertial frame is as good as another), each inertial frame corresponding to a different preferred fibration of space-time,they all shared a unique time (absolute simultaneity). In short, there was a unique breakup of 4-dimensions into (3+1). 
In special-relativistic physics, this is no longer the case: there are an infinite number of such preferred cross-sections (one for each family of parallel space-like hyperplanes in Minkowski space). Not only is the split into spaces not unique (one inertial frame is still as good as another), but now they do not even agree on a unique time slicing (the relativity of simultaneity): there is a different foliation for each preferred fibration. In short, there is a three-parameter family of ``natural'' breakups of 4-dimensions into (3+1). So, in special-relativistic physics, and quite apart from philosophical considerations, the "process" approach has much to recommend it over the ``state'' approach.
 
General relativity is an inherently four-dimensional theory of space-time-- even more so than special relativity. There is no ``natural'' breakup of space-time into spaces and times, such as the inertial frames provide in special relativity. There are no preferred timelike fibrations or spacelike foliations.\footnote{The only ``natural'' foliation would be a family of null hypersurfaces, and null hypersurface quantization has had many advocates, starting with Dirac. For a survey, see \cite{Robinson2003}.}  Any break-up of this four-dimensional structure into a (3+1) form requires the (explicit or implicit) introduction of an arbitrary ``frame of reference,''\footnote{Einstein emphasized the amorphous nature of such a frame by calling it a ``reference-mollusc'' (see \cite{Einstein1917}, cited from \cite{Einstein1961}, p. 99).}  represented geometrically by the introduction of a fibration and foliation of space-time. Then one may speak about the ``state'' of a thing on a given hyper-surface and its evolution from hyper-surface to hyper-surface of the foliation (over some ``global time''). But such a breakup is always relative to some chosen frame of reference. There are no longer any preferred breakups in generally relativity; there is always something arbitrary and artificial about the introduction of such a frame of reference. 
The process approach seems rooted in general relativistic physics, just as it is in quantum theory (see note 20). No one has put the case more strongly than Lee Smolin:
\begin{quote}
[R]elativity theory and quantum theory each... tell us-- no, better, they scream at us-- that our world is a history of processes. Motion and change are primary. Nothing is, except in a very approximate and temporary sense. How something is, or what its state is, is an illusion. It may be a useful illusion for some purposes, but if we want to think fundamentally we must not lose sight of the essential fact that it 'is' an illusion. So to speak the language of the new physics we must learn a vocabulary in which process is more important than, and prior to, stasis. \footnote{\cite{Smolin2002a}, pp. 53; my emphases on ``state'' and ``process.''}
\end{quote}

Now the canonical formalism is based on the introduction of a fibration and foliation of space-time, \footnote{All treatment of the canonical formalism mention the need for a foliation, but most do not mention the fibration, which is needed in order to understand the geometrical significance of the lapse and shift functions (see \cite{Stachel1962}, \cite{Stachel1969}).} and by its nature tends to shift attention from processes in space-time to states of things in space.\footnote{This is not meant to imply that the full four-dimensional invariance cannot be recovered in some variant of the canonical approach. For a brief discussion of ``proposals to make the canonical formulation more covariant,'' with references, see \cite{Thiemann2002}, pp. 15-16. For a summary of a different approach, see Salisbury 2003.}  Bryce DeWitt, in his final book, has put the case in the context of quantum field theory:\footnote{\cite{DeWitt2003}, vol.1, pp. v.} 
When expounding the fundamentals of quantum field theory physicists almost universally fail to apply the lessons that relativity theory taught them early in the twentieth century. Although they carry out their calculations in a covariant way, in deriving their calculational rules they seem unable to wean themselves from canonical methods and Hamiltonians, which are holdovers from the nineteenth century, and are tied to the cumbersome (3+1)-dimensional baggage of conjugate momenta, bigger-than-physical Hilbert spaces and constraints. 
This has immediate implications for a theory of quantum gravity. Whether one should be looking for quanta of space or of space-time seems to be one essential point of difference between the canonical loop quantum gravity approach and the covariant causal set approach. 
In his exposition of the canonical approach, Carlo Rovelli asserts:
 
Space-time is a temporal sequence of spaces, or a history of space. \cite{Rovelli2003}
He asks:
What are the quanta of the gravitational field? Or, since the gravitational field is the   same entity as spacetime, what are the quanta of space?\footnote{Rovelli 2004, p.18. As will be seen in \cite{Stachel2004b}, this is not the only instance of Rovelli's wavering between "space" and ``spacetime.'' There I also discussed the interpretation of the discrete spectra, in loop quantum gravity, of the operators for three-volume and two-volume and two-area on the initial hypersurface. Attributing direct physical significance to such mathematical results obtained \textit{before} solution of the Hamiltonian constraints seems to violate the general-relativitics golden rule: `` no kinematics before dynamics.''  I thank Fottini Markopoulou for helpful discussion of this point.}
The unremarked shift from ``quanta ... of spacetime'' to ``quanta of space'' is striking, but almost seems forced on Rovelli by the canonical, ``history of space'' approach. 
On the other hand, discussing causal set theory, a "process" approach to quantum gravity,\footnote{The causal set approach. to be discussed in Sections 5 and 6, does not attempt a quantization of the classical theory. Rather,  its aim is to construct a quantum theory of causal sets based on two features of classical general relativity that it takes as fundamental: 

1) the causal structure, which is replaced by a discrete causal set; and
 
2)the four-volume element, which is replaced by the quantum of process.

It must then be shown that the classical equations can be recovered from some sort of limit of  causal sets, or of an ensemble of such sets.} Fay Dowker states:
Most physicists believe that in any final theory of quantum gravity, space-time itself will be quantized and grainy in nature. .... So the smallest possible volume in four-dimensional space-time, the Planck volume, is 10-42 cubic centimetre seconds. If we assume that each of these volumes counts a single space-time quantum, this provides a direct quantification of the bulk (\cite{Dowker2003}, pp. 38). 

\subsection{Formalism and measurability}

There has always been a dialectical interplay between formalism and measurability in the development of quantum theory, first seen in the discussion about the physical interpretation of the commutation relations in non-relativistic quantum mechanics,\cite{Heisenberg1930}  and later in the similar discussion in quantum electrodynamics.\footnote{See, e.g., \cite{Bohr1933}.}  This interplay was well expressed by Bohr and Rosenfeld in their classic discussion of the measurability of the components of the electric and magnetic field:
\begin{quote}
[O]ur task will thus consist in investigating whether the complementary limitations on the measurement of field quantities, defined in this way, are in accord with the physical possibilities of measurement (Bohr and Rosenfeld 1933, cited from Rosenfeld 1979, p. 358).
\end{quote}

By ``in this way,'' they mean:  
the field quantities are no longer represented by true point functions but by functions of space-time regions, which formally correspond to the average value of the idealized field components over the region in question" (ibid.), and that delta-function commutation relations at points must be replaced by commutation relations smeared over such (finite) regions of space-time (ibid.).
 
By 1933, quantum electrodynamics had been developed to a point that enabled Bohr and Rosenfeld ``to demonstrate a complete accord'' between the formal commutation relations of the field components and the physical possibilities of their measurement. In the case of quantum gravity,  the theory is still in a state of active exploration and development, and one may hope that investigation of the possibilities of ideal measurements of the variables basic to each approach can help to clarify still unresolved issues in the physical interpretation of the formalism, and perhaps even help in choosing between various formalisms. \footnote{See \cite{Stachel2001} for further comments on this question.}
 It has been objected that a quantum theory of gravity requires a different interpretation of quantum theory, in which no external "observer" is introduced. First of all, a quantum process does not require an "observer" in any anthropomorphic sense. Once the results of a preparation and registration are recorded (see note 20), the quantum process is over, whether anyone ever looks at (``observes'') the results or not. 
Carlo Rovelli has argued well against combining the problem of quantum gravity with the problem of the interpretation of quantum mechanics:
 
I see no reason why a quantum theory of gravity should not be sought within a standard interpretation of quantum mechanics (whatever one prefers). Several arguments have been proposed to connect these two problems. A common one is that in the Copenhagen interpretation the observer must be external, but it is not possible to be external from the gravitational field. I think that this argument is wrong; if it was correct it would apply to the Maxwell field as well. We can consistently use the Copenhagen interpretation to describe the interaction between a macroscopic classical apparatus and a quantum-gravitational phenomenon happening, say, in a small region of (macroscopic) space-time. The fact that the notion of spacetime breaks down at short scale within this region does not prevent us from having the region interacting with an external Copenhagen observer (\cite{Rovelli2004}, pp. 268). \footnote{I disagree on one major point. There is one big difference between the Maxwell field and the gravitational field: the non-universality of the electromagnetic charge-current vector versus the universality of gravitational stress-energy tensor. Because charges occur with two signs that can neutralize each other, a charge-current distribution acting as a source of an electromagnetic field can be manipulated by matter that is electrically neutral and so not acting as a source of a further electromagnetic field; and one can shield against the effects of a charge-current distribution. Because mass comes with only one sign, all matter (including non-gravitational fields) has a stress-energy tensor, no shielding is possible, and any manipulation of matter acting as a source of gravitational field will introduce an additional stress-energy tensor as a source of gravitational field. A glance at Bohr and Rosenfeld 1933 shows how important the possibility of neutralizing the charges on test bodies is for measurement of the (averaged) components of the electric field with arbitrary accuracy, for example. This difference may well have important implications for the measurement of gravitational field quantities.}
Rovelli has discussed the physical interpretation of the canonical formalism for a field theory describing some generic field, symbolized by $F$: 
The data of a local experiment (measurements, preparation, or just assumptions) must in fact refer to the state of the system on the entire boundary of a finite spacetime region. The field theoretical space ... is therefore the space of surfaces $S$ [``where $S$ is a 3d surface bounding a finite spacetime region''] and field configurations f [of the $F$-field] on $S$. Quantum dynamics can be expressed in terms of an [probability] amplitude $W[S,f]$. ... Notice that the dependence of $W[S,f]$ on the geometry of $S$ codes the spacetime position of the measuring apparatus. In fact, the relative position of the components of the apparatus is determined by their physical distance and the physical time elapsed between measurements, and these data are contained in the metric of $S$. (\cite{Rovelli2004}, pp. 15-16).
 
From the ``process'' viewpoint (see Section 5a), this is an encouraging approach: what occurs in the  space-time region bounded by S constitutes a process, and an amplitude is only defined for such processes. However, as Rovelli emphasizes, this definition is broad enough to include a background-dependent theory, i.e., a theory with a fixed background space time. But his real subject of course are theories, such as general relativity, which are background-independent:
Consider a background independent theory. Diffeomorphism invariance implies immediately that $W[S,f]$ is independent from $S$. ... Therefore in gravity $W$ depends only on the boundary value of the fields. However, the fields include the gravitational field, and the gravitational field determines the spacetime geometry. Therefore the dependence of $W$ on the fields is still sufficient to code the relative distance and time separation of the components of the measuring apparatus! (ibid., pp. 16).
He summarizes the contrast between the two cases:
What is happening is that in background dependent QFT we have two kinds of measurements: the ones that determine the distances of the parts of the apparatus and the time elapsed between measurements, and the actual measurements of the fields' dynamical variables. In quantum gravity, instead, distances and time separations are on the same ground as the dynamical fields. This is the core of the general relativistic revolution, and the key for background independent QFT (Ibid., p. 16). 
This is a brilliant exposition of the nature of the difference between background dependent and background independent QFTs. But it immediately raises a number of questions, most of which I shall discuss in \cite{Stachel2004b}. Here I shall mention only one of them. 
 
Rovelli's interpretation of the canonical formalism seems based on the assumption that, in both background-dependent and background-independent QFTs, one can idealize a field measurement by confining its effects to the (three-dimensional) boundary of a (four-dimensional) space-time region; and in particular to a finite region of (three-dimensional) space, while neglecting its finite (one-dimensional) duration in time. Yet, since the pioneering work of Bohr and Rosenfeld on the measurability of the components of the electromagnetic field (Bohr and Rosenfeld 1974 [1933, 1950]), it has been known that this is not the case for the electromagnetic field in Minkowski space-time.
\begin{quote}
It is ... of essential importance that the customary description of an electric field in terms of its components at each space-time point, which characterizes classical field theory and according to which the field should be measurable by means of point charges in the sense of electron theory, is an idealization which has only restricted applicability in quantum theory. This circumstance finds its proper expression in the quantum-electromagnetic formalism, in which the field quantities are no longer represented by true point functions but by functions of space-time regions, which formally correspond to the average values of the idealized field components over the region in question. The formalism only allows the derivation of unambiguous predictions about the measurability of such region-functions ... [U]nambiguous meaning can be attached only to space-time integrals of the field components (Bohr and Rosenfeld 1933, p. 358, p. 361).
\end{quote} 

The aim of the paper was to show that these theoretical limits on measurability coincide withe the actual possibilities of (idealized)measurements of these field quantities (see Section 5b). Bohr and Rosenfeld demonstrate in some detail how test bodies occupying a finite region of space over a finite period of time– that is, test processes over finite regions of space-time– can be used to measure electromagnetic field averages over these regions.\footnote{One may see in this work the germ of the algebraic approach to quantum field theory, which also is based on attaching primary significance to operators defined over finite regions of space-time. The primary physical interpretation of the [quantum field] theory is given in terms of local operations, not in terms of particles. Specifically, we have used the basic fields to associate to each open region $O$ in space-time an algebra $A(O)$. Of operators in Hilbert spoace, the algebra generated by all $F(f)$, the fields 'smeared out' with test functions f having their support in the region $O$. We have interpreted the elements of $A(O)$ as representing physical operations performable within $O$ and we have seen that this interpretation tells us how to compute collision cross sections once the correspondence $O$ ? $A(O)$ is known. This suggests that the net of algebras $A$, i.e, the correspondence [given above] constitutes the intrinsic mathematical description of the theory. The mentioned physical interpretation establishes the tie between space-time and events. The role of ``fields'' is only to provide a coordinatization of this net of algebras" (\cite{Haag1996}, pp. 105). This approach of course extensively utilizes the existence and properties of the background Minkowski space-time. In the section on causal sets, I shall cite Haag's speculations about how it might be generalized in the absence of such a background space-time.} 
Schweber (1961) summarizes the point nicely:
\begin{quote}
In fact, Bohr and Rosenfeld in their classic papers on the question of the measurability of electromagnetic fields have already shown that only averages over small volumes of space-time of the field operators are measurable .... This is because of the finite size of the classical measuring apparatus and the finite times necessary to determine forces through their effects on macroscopic test bodies (pp. 421-422).
\end{quote}

Dosch, M\"{u}ller and Sieroka in \cite{Dosch2004} generalizes the point to all quantum field theories:
\begin{quote}
The quantum fields, in terms of which the theory is constructed, are operators that depend on space-time … This dependence on space-time however, shows the behavior of a generalized function or distribution. Therefore, a well-defined operator cannot be related to a definite space-time point x, but only to a space-time domain of finite extent (pp. 4).
\end{quote}

Sorkin in \cite{Sorkin1993} contrasts non-relativistic quantum mechanics and quantum field theory: 
 \begin{quote}
Now in non-relativistic quantum mechanics, measurements are idealized as occurring at a single moment of time. Correspondingly the interpretive rules for quantum field theory are often stated in terms of ideal measurements which take place on Cauchy hypersurfaces. However, in the interests of dealing with well-defined operators, one usually thickens the hypersurface, and in fact the most general formulations of quantum field theory assume that there corresponds to any open region of spacetime an algebra of observables which– presumably– can be measured by procedures occurring entirely within that region. \footnote{See the previous note}
\end{quote}

The question of whether measurements should be associated with (three-dimensional) things or (four-dimensional) processes recurs in loop quantum gravity, and will be discussed in 
\cite{Stachel2004b}. 

\section{Canonical quantization (loop quantum gravity).}

From the point of view of the principle of maximal permutability, the basic problem of the canonical formalism is that, by introducing a fibration and a foliation, it violates the principle. It reduces the full, four-dimensional diffeomorphism group to the subgroup that preserves a fibration and foliation of space-time\footnote{I do not mean by this that this subgroup preserves a particular fibration and foliation, but that each diffeomorphism in the subgroup takes the points of one fibration and foliation into the corresponding points of another fibration and foliation. Invariance under spatial diffeomorphisms on each hypersurface of the foliation is obviously included.},  thus losing manifest four-diffeomorphism invariance; and there certainly is a price to pay for this. Since the canonical formalism preserves spatial diffeomorphism invariance within each hypersurface of the foliation, it is perhaps not too surprising that many of its problems have to do with time and dynamics. \footnote{Speaking about loop quantum gravity, Perez states, ``The dynamics is governed by the quantum Hamiltonian constraint. Even when this operator is rigorously defined it is technically difficult to characterize its solution space. This is partly because the (3+1) decomposition of spacetime (necessary in the canonical formulation) breaks the manifest 4-diffeomorphism invariance of the theory making awkward the analysis of the dynamics'' (\cite{Perez2003}, pp. R 45).} 
\begin{quote}
The difficulty in dealing with the scalar constraint [i.e., the Hamiltonian] is not surprising. The vector constraint, generating space diffeomorphisms, and the scalar constraint, generating time reparametrizations, arise from the underlying $4$-diffeomorphism invariance of gravity. In the canonical formulation the $3+1$ splitting breaks the manifest four-dimensional symmetry. The price paid is the complexity of the time reparametrization constraint $S$ (i.e.,  the Hamiltonian, \cite{Perez2003}, pp.R 49).
\end{quote}

It is also clear that, by its nature, the canonical formalism favors a ``state of things'' over a ``process'' approach. to quantum gravity. Whether this is a drawback; and if so, how it can be overcome are questions discussed in \cite{Stachel2004b}, together with a number of other questions raised by the modern canonical formulation of general relativity (``loop quantum gravity''). Here I shall only comment on a question raised in the previous section: the interplay between formalism and measurement.
Ashtekar and Lewandowski 2004 describes the strategy adopted in loop quantum gravity:
To pass to the quantum theory [of a fully constrained theory in its phase space formulation ]\footnote{For detailed discussions of what is called ``refined algebraic quantization'' of a system with first class constraints, see \cite{Thiemann2002}, pp. 22-25, and \footnote{Thiemann 2001}, Section II.7, pp. 280-284.}, one can use one of the two standard approaches: i) find the reduced phase space of the theory representing 'true degrees of freedom' thereby eliminating the constraints classically and then construct a quantum version of the resulting unconstrained theory; \footnote{In this paper, I shall not discuss alternative i). Yet the possibilities of this alternative approach to quantization should be kept in mind. In particular, the (2+2) formalism (for a recent summary with references to earlier literature, see d'Inverno 2003) offers a possible classical starting point for such an approach, in which the two degrees of freedom of the inertio-gravitational field are given a direct geometrical interpretation. } or ii) first construct quantum kinematics for the full phase space ignoring the constraints, then find quantum operators corresponding to the constraints and finally solve the quantum constraints to obtain the physical states.\footnote{For quantum kinematics, see, e.g., \footnote{Thiemann2001}, Section I.3.}  Loop quantum gravity follows the second avenue ... (pp. 51).
 
 Note that, in this approach, the commutation relations are simply postulated. For the physical interpretation and validation of the formalism, the configuration variables and their conjugate momenta should be given some interpretation in terms of procedures for their individual measurement; and– a crucial second step-- the limitations on the joint definability of pairs of such variables implied by the postulated commutation relations should coincide with the limitations on their joint measurability by these procedures. 
The question of the link between measurability and formalism seems especially acute in view of the claim by Ashtekar and Lewandowsi that there is essentially only one possible representation of the algebraic formalism, and the suggestion by Thiemann that this mathematical uniqueness may have been achieved at too high a physical price.\footnote{The main point of the following discussion should be clear even to readers unfamiliar with the canonical formalism: It is claimed that, given the techniques that make the new canonical quantization possible, there is a just one possible operator representation of the algebra of the basic canonical variables. If correct, then it is crucial to make sure that the possibilities for actual measurement of these variables reaches, but does not exceed, the limits set by this algebra.} As explained in detail in \cite{Stachel2004b}, in loop quantum gravity the holonomies associated with a certain three-connection are taken as the "configuration" variables, with a corresponding set of so-called electric-flux "momentum" variables, which obey a certain algebra.
The main problem is that of finding the physically appropriate representation of the holonomy-flux algebra (\cite{Ashtekar2004}, pp. 41).
They succeed in constructing one, and state:
 
Remarkably enough, uniqueness theorems have been established: there is a precise sense in which this is the only diffeomorphism invariant, cyclic representation of the kinematical quantum algebra ... Thus, the quantum geometry framework is surprisingly tight. ... there is a unique, diffeomorphism invariant representation ... (ibid., p. 41, my emphasis). \footnote{Put more technically: As we saw, quantum connection-dynamics is very 'tight'; once we choose the holonomies $A(e)$ and the 'electric fluxes' $P(S,f)$ as the basic variables, there is  essentially no freedom in the independent quantization (ibid.).}
One might have expected that, as in ordinary quantum mechanics, there is a complementary representation, based on the ``momentum'' variables conjugate to the configuration ones (these are the 'electric fluxes' mentioned above), which essentially constitute the three-metric. But these variables, suitably smeared, do not commute;\footnote{As a consequence, and even more significant for physical measurements, "area operators $AS$ and $AS'$ fail to commute if the surfaces $S$ and $S'$ intersect. ... Thus, the assertion that ``the spin network basis diagonalizes all geometrical operators' that one sometimes finds in the literature is incorrect'' (ibid., p. 46).}  consequently, 
[t]his result brings out a fundamental tension between connection-dynamics and geometrodynamics. ...[T]he metric representation does not exist (ibid., p. 46).

This uniqueness claim is a striking indeed. In Lorentz-invariant quantum field theories, there are unitarily  inequivalent representations of the basic algebra, which led to the algebraic approach to quantum field theory that de-emphasizes the role of the representations.\footnote{See \cite{Haag1996} for a full account}  But in loop quantum gravity,
[t]hese results seem to suggest that, for background independent theories, the full generality of the algebraic approach may be unnecessary (ibid., p. 41). \footnote{While the full generality of the algebraic approach might not be needed, it might still be worthwhile to investigate the possibility of a four-dimensional algebra that reduces to the holonomy-flux algebra on a space-like hypersurface. If this four-dimensional algebra could be related to physical measurements, this might provide another approach to the question, raised above, of physically justifying the holonomy-flux algebra instead of simply postulating it. See note 54 for Haag's comments on the possibility of a net of algebras on a partially ordered set.}
 
But some caution is advisable here; this result is based on the requirement of spatial diffeomorphism invariance. But, as Thiemann notes,
\begin{quote}
[t]here are, however, doubts on physical grounds whether one should insist on spatial diffeomorphism invariant representation because smooth and even analytic structure of [the three-manifold] which is encoded in the spatial diffeomorphism group should not play a fundamental role at short scales if Planck scale physics is fundamentally discrete.  In fact, as we shall see later, Q[uantum] G[eneral] R[elativity] predicts a discrete Planck scale structure and therefore the fact that we started with analytic data and ended up with discrete (discontinuous) spectra of operators looks awkward. Therefore ... we should keep in mind that other representations are possibly better suited in the final picture ... (\cite{Thiemann2002}, pp. 40; see also the quotation from \cite{Thiemann2001} at the end of Section 2). 
\end{quote}
   
Are there other representations once one drops the demand for diffeomorphism invariance? If there are, will they be inequivalent (as in quantum field theory) or equivalent (as in non-relativistic quantum mechanics) to the Ashtekar-Lewandowski representation? In view of the physical importance of the answers to these technical questions, one would like to be certain that the formal success of this possibly-unique quantization is accompanied by an equally successful physical interpretation of it. Ashtekar and Lewandowski are far from claiming that this has been accomplished. In discussing quantum dynamics, they note that:
 \begin{quote}
the requirement of diffeomorphism invariance picks out a unique representation of the algebra generated by holonomies and electric fluxes. Therefore we have a single arena for background independent theories of connections and a natural strategy for implementing dynamics provided, of course, this mathematically natural, kinematic algebra is also 'physically correct' (ibid., p. 51).
\end{quote}

Once these ``kinematic'' steps have been taken, there still remains the final ``dynamical'' step:
We come now to the ``Holy Grail'' of Canonical Quantum General Relativity, the definition of the Hamiltonian constraint. .... 
\begin{quote}
[T]he implementation of the correct quantum dynamics is not yet completed and one of the most active research directions at the moment (\cite{Theimann2001}, pp. 150).
\end{quote}

Another question can be raised in this connection. Might it be possible to go beyond simply postulating the three-dimensional commutation relations in loop quantum gravity, which after all do not take us beyond one spatial hypersurface?  Various spin foam models (see \cite{Perez 2003}) are candidates for a four-dimensional, dynamical version of the canonical approach. So it might be possible to introduce four-dimensional commutation relations in such models (insofar as these models can be freed from their origins in, and close ties to, the canonical formalism-- see \cite{Stachel2004b}) and show that the three-dimensional canonical commutation relations postulated on a space-like hypersurface can be derived from them.\footnote{``[I]t would be very interesting to reconstruct in detail the hamiltonian Hilbert space, as well as kinematical and dynamical operators of the loop theory, starting from the covariant spinfoam definition of the theory. At present [this problem is not] under complete control'' (\cite{Rovelli2004}, pp. 362).}  Since the presence of a causal structure seems necessary for the formulation of four-dimensional commutation relations, causal spin foam models might be good candidates for such an approach (see \cite{Stachel2004b}).

\section{The causal set (causet) approach}

I should like to draw attention to an early work on this subject, which I have not seen cited in the recent literature on causal sets \cite{Kronheimer1967}.

This approach would seem to have already adopted much of the viewpoint suggested here. \footnote{For a popular survey, see \cite{Dowker2003}. For a more technical survey, see 
\cite{Sorkin2003}}  Space-time points are replaced by quanta of process, elements of four-volume of order (LP)4, and these are the basic entities in this approach.\footnote{Rudolf Haag's comments on how the algebraic approach to field theory might be modified in the absence of a bckground space-time bear a close resemblance to the causal set approach. "In a minimal adaptation of the algebraic approach and the locality principle one could keep the idea of a net of algebras which, however, should be labeled now by the elements of a partially ordered set L (instead of regions in R4). L could be atomic, with minimal elements (atoms) replacing microcells in space-time " (\cite{Haag1996}, pp. 348).} They are connected to each other by a causal (partial) ordering relation. The causal ordering relation enables us to define a causal past and a causal future for each element of the causal set, forming the discrete analogues of the forward and backward light cones of a point, which define the  classical causal structure of a space-time. The number of quanta of process in any given four-dimensional process determines its space-time "volume." Together, they provide the causal-set analogue of the four-dimensional metric tensor. 
After introducing the idea of a ``labeled causet,'' in  which each element of the causet is labeled by the sequence in which it is introduced, Sorkin et al comment: \footnote{See \cite{Brightwell2002}, pp. 8.}
 
After all, labels in this discrete setting are the analogs of coordinates in the continuum, and the first lesson of general relativity is precisely that such arbitrary identifiers must be regarded as physically meaningless: the elements of spacetime - or of the causet – have individuality only to the extent that they acquire it from the pattern of their relations to the other elements. It is therefore natural to introduce a principle of "discrete general covariance" according to which ``the labels are physically meaningless''. 
But why have labels at all then? For causets, the reason is that we don't know otherwise how to formulate the idea of sequential growth, or the condition thereon of Bell causality, which plays a crucial role in deriving the dynamics. Ideally perhaps, one would formulate the theory so that labels never entered, but so far, no one knows how to do ' this -- anymore than one knows how to formulate general relativity without introducing extra gauge degrees of freedom that then have to be canceled against the diffeomorphism invariance. 
Given the dynamics as we can formulate it, discrete general covariance plays a double       role. On one hand it serves to limit the possible choices of the transition probabilities in    such a way that the labels drop out of certain ``net probabilities'', a condition made        precise in [ref deleted]. This is meant to be the analog of requiring the gravitational       action-integral $S$ to be invariant under diffeomorphisms (whence, in virtue of the further    assumption of locality, it must be the integral of a local scalar concomitant of the           metric). On the other hand, general covariance limits the questions one can                    meaningfully ask about the causet (cf. Einstein's ``hole argument''). It is this second         limitation that is related to the ``problem of time'', and it is only this aspect of discrete   general covariance that I am addressing in the present talk. 
 
I think there is a certain confusion here between the causet analogues of active and passive diffeomorphisms.  Sorkin rightly emphasizes that ``the elements of spacetime- or of the causet – have individuality only to the extent that they acquire it from the pattern of their relations to the other elements.'' Yet he goes on to speak about invariance under a re-labeling of the elements of the causet as ``discrete general covariance.'' But such a re-labeling corresponds to a coordinate transformation, i.e., a  passive diffeomorphism. What is important physically is an active permutation of the elements of the causet. But clearly all the physically relevant information about the causet is contained in the network of order relations between its elements and,  as long as this is not changed, a permutation of the elements changes nothing. In my terminology, the elements of the causet have quiddity– crudely put, they are quanta of four-volume – but lack haecceity– nothing distinguishes one from the other except its position in the causet. So, in spite of Sorkin's terminology, there is no problem here. 
The chief defect of the causal set approach is that so far it is not really a quantum theory; that is, it  has not been able to take the step from transition probabilities to transition probability amplitudes, which would allow a Feynman formulation of the theory, leading to ``a 'sum over histories' quantum theory of causets'' \cite{Dowker2004}, i.e., the addition of the amplitudes for indistinguishable processes that begin with the same initial preparation and end with the same final registration. 
Perhaps connected with this problem is the circumstance that the theory is not very closely linked to classical general relativity. It simply postulates certain things - such as the discreteness of processes (i.e., four-volumes):

The number of causet elements gives the volume of the corresponding region of the approximating spacetime in Planck units \footnote{see\cite{Dowker2004}. See also the more complete discussion in \cite{Sorkin2003}, which offers several heuristic arguments.}-  that one might hope to derive from a quantum version of general relativity, for example by an appropriate quantization of the conformal factor in the metric (i.e., the determinant of the metric tensor). In the last section, I shall offer some suggestions about how this might be done.	

\section{What Structures to Quantize?}

There are a number of space-time structures that play an important role in the general theory of relativity (see \cite{Stachel2003}). The chrono-geometry is represented mathematically by a pseudo-metric tensor field on a four-dimensional manifold. The inertio-gravitational field \footnote{This is often referred to simply as the gravitational field. But it must be emphasized that the deeper meaning of the equivalence principle is that there is no frame-independent separation between inertia and gravitation. This is as true of Newton's gravitational theory, properly interpreted as a four-dimensional theory, as it is of general relativity} is represented by a symmetric affine connection on this manifold. Then there are compatibility conditions between these two structures (the covariant derivative of the metric with respect to the connection must vanish). As noted above, major technical advances in the canonical quantization program came when the inertio-gravitational connection was taken as primary rather than the chrono-geometrical metric.
It is possible to start with only the metric field and derive from it the unique symmetric connection (the Christoffel symbols) that identically satisfies the compatibility conditions. A second order Lagrangian (the densitized curvature scalar) then leads to the field equations in terms of the metric. This is the route that was first followed historically by Hilbert, and is still followed in most textbooks.
 
It is also possible to treat metric and connection as initially independent structures, and then allow the compatibility conditions between them to emerge, together with the field equations (written initially in terms of the connection), from a first order, Palatini-type variational principle. In this approach, metric and connection are in a sense "dual" to each other.
Either of these methods may be combined with a tetrad formalism for the metric, combined with one or another mathematical representation of the connection, e.g., connection one-forms, or tetrad components of the connection.
But one can go a step further and decompose the metric and affine connection themselves. If one abstracts from the four- volume-defining property of the metric, one gets the conformal structure on the manifold. Physically, this conformal structure is all that is needed to represent the causal structure of space-time. Similarly, if one abstracts from the preferred parametrization (proper length along spacelike, proper time along timelike) of the geodesics associated with an affine connection, \footnote{Alternately, and perhaps better put, if one abstracts from the concept of parallel transport its property of preservation of ratios of parallel vectors, one gets the concept of projective transport, which preserves only the direction of a vector.} one gets a projective structure on the manifold. Physically the projective structure picks out the class of preferred  paths in space-time.  \footnote{Remember the distinction between curves, which are associated with a parametrization, and paths, which are not. Since the holonomies of the affine connection are independent of the parametrization of the curves, it is possible that the holonomies of the projective connection might be more suited to loop quantum gravity.}
Compatibility conditions between the causal and projective structures can be defined, and which also guarantee the existence of a corresponding metric and compatible affine connection.\footnote{See \cite{Ehlers1972}. For some further results clarifying the interrelations between conformal, (semi)Riemannian, volume and projective geometrical structures, see \cite{Sanchez-Rodriguez2001}.} 
 
But the important point here is that, as shown in \cite{Weyl1921}, in a manifold with metric, its conformal and affine structures suffice to determine that metric (up to an overall constant factor). We can use this circumstance to our advantage in the first order, so-called Palatini variational principle for general relativity. As mentioned above, in this case metric and affine connection are independently varied, but the metrical nature of the connection follows from the variation of the connection.  But one can go further: break up the metric into its four-volume structure-determining part (essentially, the determinant of the metric) and its conformal, causal structure-determining part (with unit determinant); and break up the affine connection into its projective, preferred path-determing part (the trace-free part of the connection) and its preferred parameter-determining part ( the trace of the connection). \footnote{For discussions of projective and conformal geometry, see \cite{Schouten}, Chapter 6, pp. 287-334.} Each of these four parts may then be varied independently. 
Such a breakup is of particular interest because, as noted in the previous section, the causal set theory approach to quantum gravity is based on taking the conformal structure and the four-volume structure as the primary constituents of the classical theory,\footnote{We can say that a spacetime is its causal structure (order) plus volume information" (\cite{Dowker2004}).}  and then replacing them with discretized versions:\footnote{Ibid.}  
The causal set idea combines the twin ideas of discreteness and order to produce a structure on which a theory of quantum gravity can be based.. 
 
However, in causal set theory no attention seems to have been paid to the affine connection, and the possibility of finding quantum analogues of the traceless projective and trace parts of the connection. 
The answer to this question might lead to a link between the causal set approach and some quantized version of the dynamics of general relativity. It is easy to set up a conformal-projective version of the Palatini principle, in which the trace of the affine connection is dual to the four-volume structure, suggesting that the projective connection is dual to the conformal structure. So a covariant quantization based on this breakup might lead to a representation in which four-volume and conformal structures are the configuration variables - a sort of four-dimensional analogue of quantum geometrodynamics; with the possibility of another representation in which the projective connection and the trace factor are the configuration variables– a sort of four-dimensional analogue of the loop quantum gravity representation. 
Of course such an approach would seem to require a full, four dimensional quantization procedure rather than a canonical one. As suggested above, such an approach might provide a way to derive the fundamental quantum of process in general relativity. Quanta of (proper) three-volume and (proper) time might then be related to ``perspectival'' effects of ``viewing'' quanta of four-volume by observers in different states of relative motion through space-time.

\section{Acknowledgements}

\hspace{0.7 cm}I thank Carlo Rovelli for written comments, and Abhay Ashtekar and Thomas Thiemann for oral comments, on an earlier version of this paper. Of course, I am solely responsible for my comments on their work. 

I thank Mihaela Iftime, whose critical reading of the text led to a number of important clarifications and improvements and who prepared the LaTeX version.


\begin{thebibliography}{100}


\bibitem{Adams1979}\textbf{Adams, Robert M.} (1979),
\textit{Primitive Thisness and Primitive Identity}, The Journal of Philosophy {\bf 76}, pp.5-26

\bibitem{Ashtekar1987}\textbf{Ashtekar, A.} (1987), 
\textit{Asymptotic Quantization} Naples: Bibliopolis

\bibitem{Ashtekar1999}\textbf{Ashtekar, A.} (1999), 
\textit{Discussions},  Quantum Field Theory of Geometry in Cao, pp. 203-206.

\bibitem{Ashtekar2004}\textbf{Ashtekar, A. and Lewandowski,J} (2004), 
\textit{Background Independent Quantum Gravity: A Status Report} 
arXiv:gr-qc/0404018v1, 5 April 2004

\bibitem{Auyang1995}\textbf{Auyang, Sunny Y.} (1995), 
\textit{How is Quantum Field Theory Possible?}
 New York/Oxford, Oxford University Press. 

\bibitem{Baez2002}\textbf{Baez, John C., Christensen, J. Daniel; Halford, Thomas R.; and Tsang, David C.} (2002), 
\textit{Spin foam models of Riemannian quantum gravity} 
Classical and Quantum Gravity {\bf 19+},pp. 4627-4648.
 
\bibitem{Bohr1933}\textbf{Bohr, Niels and Rosenfeld, Leon} (1933), 
\textit{Zur Frage der Messbarkeit der elektromagnetischen Feldgrössen},
 Mat-fys. Medd. Dan. Vid. Selsk 12, no. {\bf 8}, cited from the English translation, 
\textit{On the Question of the Measurability of the Electromagnetic Field Quantities}
in Robert S. Cohen and John Stachel, eds., Selected Papers of Leon Rosenfeld. Dordrecht/Boston/London: D. Reidel 1978, pp.357-400.

\bibitem{Brightwell2002}\textbf{Brightwell, Graham; Dowker, H. Fay; García; Henson, Joe; Sorkin, Rafael D.} (2002), 
\textit{General Covariance and the 'Problem of Time' in a Discrete Cosmology} ArXiv:gr-qc/0202097,to appear in the proceedings of the Alternative Natural Philosophy Association meeting, August 16-21, Cambridge, England


\bibitem{Burgess2004}\textbf{Burgess, Cliff P.} (2004), 
\textit{Quantum Gravity in Everyday Life: General Relativity as an Effective Field Theory}\\ Available online: www.livingreviews.org/lrr-2004-5

\bibitem{Cao1999}\textbf{Cao, Tian Yu, ed.} (1999), 
\textit{Conceptual Foundations of Quantum Field Theory} 
Cambridge University Press.

\bibitem{DeWitt2003}\textbf{DeWitt, Bryce} (2003),
\textit{The Global Approach to Quantum Field Theory} {\bf 2} vols, 
Oxford: Clarendon  Press.

\bibitem{D'Inverno}, \textbf{D' Inverno, Ray A.} (2003), 
\textit{DSS 2+2} 
in Abhay Ashtekat et al, eds., Revisiting the Foundations of Relativistic Physics/ Festschrift in Honor of John Stachel. Dordrecht/Boston/London: Kluwer Academic, pp. 317-347.


\bibitem{Donoghue1995}\textbf{ Donoghue, John F.} (1995), 
\textit{Introduction to the Effective Field Theory Description of Gravity}
arXiv:gr-qc/9512024 v1 11 Dec 1995

\bibitem{Dosch2004}\textbf{Dosch, Hans Günther, Volkard F. Müller and Norman Serioka} (2004), 
\textit{Quantum Field Theory, Its Concepts Viewed from a Semiotic Perspective}, 
philsci-archive.pitt.edu/archive/ 00001624/01/ms

\bibitem{Dowker2003}\textbf{Dowker, F.} (2003), 
\textit{Real Time},
New Scientist 180, pp. 36-39.


\bibitem{Dowker2004}\textbf{Dowker, F.}(2004), 
\textit{Causal Sets as the Deep Structure of Spacetime} 
transparencies for a lecture.
 

\bibitem{Ehlers1972}\textbf{Ehlers, Jürgen, Pirani, Felix, and Schild, Alfred} (1972), 
\textit{The Geometry of Free Fall and Light Propagation},
 in L. O'Raifertaigh, ed., General Relativity/ Papers in Honor of J. L. Synge. Oxford: Clarendon Press, pp. 63-84.

\bibitem{Ellis2002}\textbf{Ellis, George F. R.}, (2002)
\textit{The Universe Around Us: AN Integrative View of Science and Cosmology.}
www.mth.uct.ac.za/~ellis/cosa.html

\bibitem{Einstein1917}\textbf{Einstein, Albert} (1917), 
\textit{Uber die spezielle und die allgemeine Relativitatstheorie(Gemeinverstänlich)}, Braunschweig: Friedr. Vieweg \& Sohn

\bibitem{Einstein1961}\textbf{Einstein, Albert} (1961), 
\textit{Relativity/The Special and General Theory}, 
15th ed. transl. by Robert W. Lawson,. New York: Crown Publishers.

\bibitem{Filk2001}\textbf{Filk, Thomas} (2001), 
\textit{Proper time and Minkowski structure on causal graphs},
 Classical and Quantum Gravity {\bf 18}, pp. 2785-2795.

\bibitem{Greene2004}\textbf{Greene, Brian} (2004), 
\textit{The Fabric of the Cosmos/ Space, Time and the Texture of Reality}. 
New York: Alfred A. Knopf.

\bibitem{Haag1996}\textbf{Haag, Rudolf} (1996), 
\textit{Local Quantum Physics/ Fields, Particles, Algebras}.
2nd ed. Berlin/Heidelberg/New York: Springer Verlag.

\bibitem{Heisenberg1930}\textbf{Heisenberg, Werner} (1930), 
\textit{The Physical Principles of the Quantum Theory}.
Chicago: University of Chicago Press. Cited from the Dover Press reprint.

\bibitem{Wald}\textbf{Hollands, Stefan and Wald, Robert M.} (2002), 
\textit{Existence of Local Covariant Time Ordered Products of Quantum Fields in Curved Spacetime},  Communications in Mathematical Physics {\bf 231}, pp. 309-345.

\bibitem{Krause2004}\textbf{Krause, D\'{e}cio} (2004), 
\textit{Structures and Structural Realism},
 Available on the website http://philsci archive.pitt.edu/archive/00001558/01/OnticReal2004.pdf
 
\bibitem{Kronheimer1967}\textbf{Kronheimer, E. H. and Penrose, Roger} (1967), 
\textit{On the Structure of Causal Spaces},
Proceedings of the Cambridge Philosophical Society {\bf 63}, pp. 481-501.

\bibitem{Ladyman1998}\textbf{Ladyman, James} (1998), 
\textit{What is Structural Realism?},  
Studies in the History and Philosophy of Science {\bf 29}, pp. 49-424.

\bibitem{Mallios1998}\textbf{Mallios, A} (1998), 
\textit{On an Axiomatic Treatment of Differential Geometry via Vector Sheaves. Applications}, Mathematica Japonica {\bf 48}, pp. 93-180.

\bibitem{Mallios2004}\textbf{Mallios, A} (2004), 
\textit{Gauge Theories from the Point of View of Abstract Differential Geometry},
{\bf 2} vols., to appear.

\bibitem{Mallios2003}\textbf{Mallios, A. and Raptis, Ionannis},(2003) 
\textit{Finitary, Causal and Quantal Vacuum Einstein Gravity}, 
International Journal of Theoretical Physics {\bf 42}, pp. 1479-1619.

\bibitem{Markopoulou2002}\textbf{Markopoulou, Fotini} (2002),  
\textit{Planck-scale models of the Universe},
ArXiv:gr-qc/0210086v1.


\bibitem{Markopoulou2004}\textbf{Markopoulou, Fotini}(2004), 
\textit{Particles in a Spin Foam},  
transparencies for a talk at the QI/QG Workshop, Perimeter Institute, February 2004. \\
Available online: 
www.perimeterinstitute.ca/activities/ scientific/cws/feb2004workshop/fotini.pdf

\bibitem{Markopoulou1997}\textbf{Markopoulou and Smolin} (1997), 
\textit{Causal evolution of spin networks} 
arXiv:gr-qc/9702025 v1 


\bibitem{Perez2003}\textbf{Perez, Alejandro} (2003), 
\textit{Spin Foam Models for Quantum Gravity} 
Classical and Quantum Gravity {\bf 20}, pp. R43-R104.


\bibitem{Rideout2000}\textbf{Rideout, D. P. and Sorkin, Rafael D} (2000), 
\textit{Classical sequential growth dynamics for causal sets}, Physical Review D {\bf 61}, 024002.


\bibitem{Rovelli1999}\textbf{Rovelli, Carlo} (1999), 
\textit{Localization in quantum field theory: how much of QFT is compatible with what we know about space-time?} in Cao 1999, pp. 207-230.

\bibitem{Rovelli2003}\textbf{Rovelli, Carlo} (2003), 
\textit{Loop quantum gravity}, Physics World, November 2003: 37-42.

\bibitem{Rovelli2004}\textbf{Rovelli, Carlo} (2004), 
\textit{??} Quantum Gravity. Cambridge University Press.

\bibitem{Runes1962}\textbf{Runes, Dagobert D., ed.} (1962), 
\textit{Dictionary of Philosophy},  Totowa, New Jersey: Littlefield, Adams \& Co.


\bibitem{Salisbury2003}\textbf{Salisbury, Don C.} (2003)
\textit{Gauge Fixing and Observable in General Relativity}
Availabe online: arXiv:gr-qc/0310095 v1


\bibitem{Samuel2000}\textbf{Samuel, J.} (2000), 
\textit{Is Barbero's Hamiltonian formulation a Gauge Theory of Lorentzian Gravity?} 
Classical and Quantum Gravity {\bf 17}, pp. L141-L148.


\bibitem{Sanchez-Rodriguez2001}\textbf{Sanchez-Rodriguez, I.} (2001), 
\textit{Intersection of G-structures of first or second order},
 Differential Geometry and Its Applications; Proc. Conf. Opava (Czech Republic), August 27-31, 2001. Opava: Silesian University, pp. 135-140.

\bibitem{Saunders2003}\textbf{Saunders, Simon} (2003), 
\textit{Structural Realism, Again}, in Symons 2003, pp. 127-133.

\bibitem{Schouten}\textbf{Schouten, J. A.} (1954), 
\textit{Ricci-Calculus}, 2nd ed. Berlin/Göttingen/Heidelberg: Srpinger-Verlag.

\bibitem{Smolin2002a}\textbf{Smolin, Lee} (2002), 
\textit{Three Roads to Quantum Gravity} Lymington, Hants., Basic Books.

\bibitem{Smolin2002b}\textbf{Smolin, Lee}(2002), 
\textit{Technical Summary of Loop Quantum Gravity}
 Available online: www.qgravity.org/loop/

\bibitem{Sorkin1993}\textbf{Sorkin, Rafael D.} (1993), 
\textit{Impossible Measurements on Quantum Fields}, in B. L. Hu and T. A. Jacobson, eds., Directions in General Relativity, vol. {\bf 2}, Papers in Honor of Dieter Brill. Cambridge: Cambridge University Press, pp. 293-305. 
 
 
\bibitem{Sorkin1997}\textbf{Sorkin, Rafael D.}(1997), 
\textit{Forks in the Road, on the Way to Quantum Gravity},
International Journal of Theoretical Physics {\bf 36}, pp.2757-2781

\bibitem{Sorkin2003}\textbf{Sorkin, Rafael D.} (2003), 
\textit{Causal Sets: Discrete Geometry},
 Available online: ArXiv:gr-qc/0309009 v1 1 Sep 2003.

\bibitem{Stachel1962}\textbf{Stachel, John} (1962), 
\textit{Lie Derivatives and the Cauchy Problem in Generalized Electrodynamics and General Relativity}, Ph. D. Dissertation, Stevens Institute of Technology.


\bibitem{Stachel1969}\textbf{Stachel, John} (1969), 
\textit{Covariant Formulation of the Cauchy Problem in Generalized Electrodynamics and General Relativity}, Acta Physica Polonica {\bf 35},pp. 689-709.

\bibitem{Stachel1986}\textbf{Stachel, John}  (1986), 
\textit{Do Quanta Need a New Logic?}
in Robert G. Colodny, ed., From Quarks to Quasars/Philosophical Problems of Modern Physics.  Pittsburgh: University of Pittsburgh Press, pp. 229-347. 


\bibitem{Stachel1993}\textbf{Stachel, John}(1993), 
\textit{The Meaning of General Covariance: The Hole Story}, in John Earman, et al., eds., Philosophical Problems of the Internal and External Worlds 
Pittsburgh: University of Pittsburgh Press/Konstanz: Universitätsverlag, pp.129-160. 


\bibitem{Stachel1997}\textbf{Stachel, John} (1997), 
\textit{Feynman Paths and Quantum Entanglement: Is There Any More to the Mystery?}in Robert S. Cohen, Michael Horne and John Stachel, eds., Potentiality, Entanglement and Passion-at-a-Distance/ Quantum Mechanical Studies for Abner Shimony, vol. {\bf 2},  Dordrecht/Boston/London, Kluwer Academic Publishers, pp. 244-256.


\bibitem{Stachel2001}\textbf{Stachel, John} (2001),
\textit{Some Measurement Problems in Quantum Gravity}, 
unpublished text based on lectures given at the Relativity Center, Pennsylvania State University.
 
\bibitem{Stachel2002}\textbf{Stachel, John}(2002),
\textit{The Relations Between Things' versus 'The Things Between Relations': The Deeper Meaning of the Hole Argument}
in David B. Malament, ed., Reading Natural Philosophy/ Essays in the History and Philosophy of Science and Mathematics. Chicago and LaSalle, IL, Open Court, pp. 231-266. 


\bibitem{Stachel2003}\textbf{Stachel, John}(2003), 
\textit{A Brief History of Space-Time} 
in Ignazio Ciufolini, Daniele Dominic, and Luca Lusanna, eds.,  2001: A Relativistic Spacetime Odyssey/Experiments and Theoretical Viewpoints on General Relativity and Quantum Gravity: Proceedings of the 25th Johns Hopkins Workshop on Current Problems in Particle Theory. Singapore, World Scientific, pp. 15-34.


\bibitem{Stachel2004a}\textbf{Stachel, John} (2004a), 
\textit{Structural Realism and Contextual Individuality}
in Yemima Ben-Menahem, ed. Hilary Putnam. Cambridge University Press.


\bibitem{Stachel2004b}\textbf{Stachel, John}(2005), 
\textit{Some Problems of Loop Quantum Gravity}, preprint


\bibitem{Stachel2006a}\textbf{Stachel, John}(2006), 
\textit{The Story of Newstein, or Is Gravity Just Another Pretty Force ?} 
To appear in Jurgen Renn and Matthias  Schimmel, eds The Genesis if General Relativity: Documents and Interpretations, vol.{\bf 3}

\bibitem{Stachel2006b}\textbf{Stachel, John}(2006), 
\textit{Alternative Approaches to General Relativity}, Dodrecht: Kluwer 


\bibitem{Stachel2005}\textbf{Stachel, John and Iftime, Mihaela}(2005), 
\textit{Fibered Manifolds, Natural Bundles, Structured Sets, G-Spaces and All That: The Hole Story From Space-Time to Elementary Particles}, preprint, gr-qc/0505138 v2
to appear in John Stachel, \textit{Going Crtical}, vol {\bf 2}, \textit{The Practice of Relativity}, Dordercht Kluwer

\bibitem{Symons2003}\textbf{Symons, John, ed.} (2003),  
\textit{Symposium on Structural Realism and Quantum Field Theory},
Synthese 136, No.{\bf  1}.

\bibitem{Teller1995}\textbf{Teller, Paul} (1995), 
\textit{An Interpretative Introduction to Quantum Field Theory.} Princeton, N.J., 
Princeton University Press. 

\bibitem{Teller1998}\textbf{Teller, Paul} (1998) 
\textit{Quantum Mechanics and Haecceities,} 
in Elena Castellani, ed., Interpreting Bodies/Classical and Quantum Objects in Modern Physics. Princeton University Press, 
pp. 114-141.


\bibitem{Thiemann2001}\textbf{Thiemann, Thomas} (2001), 
\textit{Introduction to Modern Canonical Quantum General Relativity,}
arXiv:gr-qc/0110034v1, 5 October 2001.

\bibitem{Thiemann2002}\textbf{Thiemann, Thomas} (2002), 
\textit{Lectures on Loop Quantum Gravity,}
arXiv:gr-qc/0210094v1, 28 October 2002.

\bibitem{Thiemann2004}\textbf{Thiemann, Thomas} (2004), 
\textit{The LQG-String: Loop Quantum Gravity Quantization of String Theory I. Flat 
Target Space} arXiv:hep-th/0401172 v1, 23 January 2004.

\bibitem{Wald1994}\textbf{Wald, Robert M.} (1994), 
\textit{Quantum Field Theory in Curved Spacetime and Black Hole Thermodynamics}
Chicago and London: University of Chicago Press.


\bibitem{Weyl1921}\textbf{Weyl, Hermann} (1921), 
\textit{Zur Infinitesimalgeometrie: Einordnung der projektiven und 
konformen Auffassung,} 
Nachrichten der Königlichen Gesselschaft der Wissenschaften 
zu G$\ddot{o}$ttingen, Mathematisch-physikalische Klasse,pp. 99-112.  

\bibitem{Weyl1922}\textbf{Weyl, Hermann}  (1922), 
\textit{Zur Infinitesimalgeometrie: p-dimensional Fl$\ddot{a}$che im n-dimensionalen Raum,}
Mathematische Zeitschrift {\bf 12}, pp. 154-160.

\bibitem{Wuthrich2003}\textbf{W\"{u}thrich, Christian} (2003), 
\textit{Quantum Gravity and the 3D vs. 4D Controversy.} 
Available online:http://alcor.concordia.ca/~scol/seminars/conference/Wuthrich.pdf                  

\end{thebibliography}
\end{document}